# Quantum Geometric Engineering of Dual Hall Effects in 2D Antiferromagnetic Bilayers via Interlayer Magnetic Coupling

Zhenning Sun[1], Tao Wang[1], Hao Jin[1]*, Xinru Li[2], Yadong Wei[1]*, Jian Wang[1,3]

[1]*College of Physics and Optoelectronic Engineering, Shenzhen University, Shenzhen 518060, P. R. China*

[2]*School of Physics, State Key Laboratory of Crystal Materials, Shandong University, Shandanan Street 27, Jinan 250100, China*

[3]*Department of Physics, The University of Hong Kong, Pokfulam Road, Hong Kong 999077, China*

**ABSTRACT:** The interplay between quantum geometry and magnetic order offers a novel strategy for designing next-generation nanodevices. Here, we demonstrate that interlayer magnetic coupling in two-dimensional (2D) CoPSe$_3$ bilayers enables precise control over quantum geometric mechanisms, unlocking dual intrinsic Hall effects. Our first-principles calculations reveal that the altermagnetic (AM) phase exhibits a giant anisotropic anomalous Hall effect (AHE) ($\sigma_{xy} \approx 46$ S/cm) driven by Berry curvature localized at generic $k$-points, while the $\mathcal{PT}$-symmetric antiferromagnetic (AFM) phase hosts an intrinsic second-order nonlinear anomalous Hall effect (NAHE) ($\chi_{xyy} \approx 160$ μS/V) originating from quantum metric accumulation at high-symmetry $k$-points. By tuning interlayer magnetic couplings, we achieve reversible switching between these phases, leveraging their distinct band structures and symmetry constraints. The Néel-vector-dependent AHE in the AM phase and the symmetry-protected NAHE in the AFM phase highlight quantum geometry as a versatile tool for manipulating transport properties. Our work establishes 2D antiferromagnets as a promising platform for multifunctional device architectures, bridging linear and nonlinear magnetoelectric responses through tailored quantum geometric engineering.

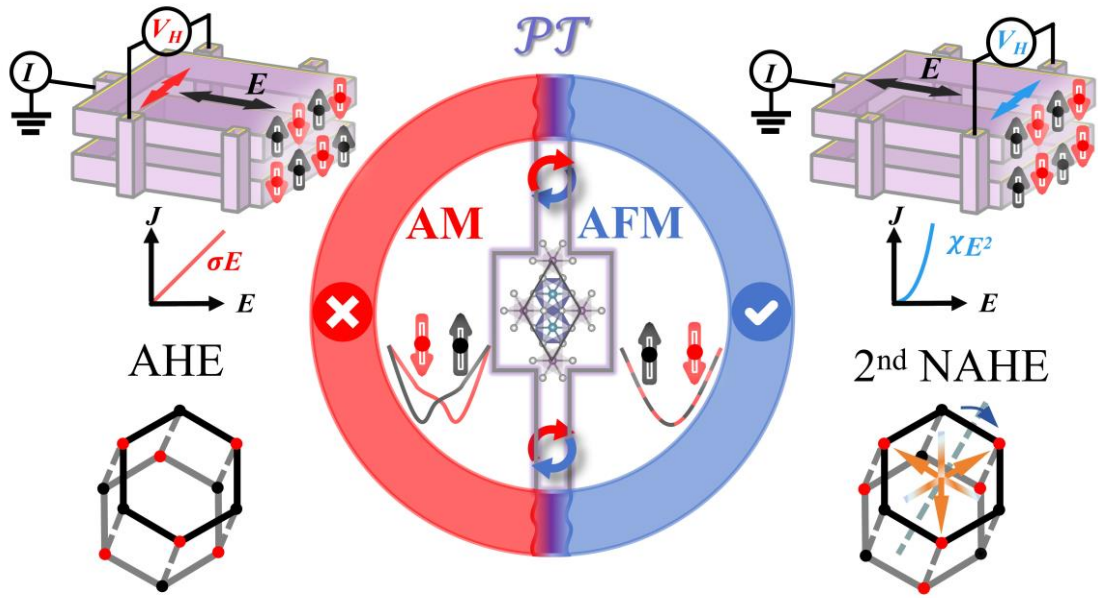

Quantum geometric tensor, which characterizes the geometric properties of electronic wavefunctions in momentum space, plays a fundamental role in determining the transport properties of materials, particularly Hall effects.[1-5] This quantum geometric tensor has both real and imaginary components,[6-14] each contributing uniquely to the Hall response. The imaginary part, known as the Berry curvature, is central to the intrinsic anomalous Hall effect (AHE), a phenomenon that is independent of the relaxation time $\tau$ and is typically observed in ferromagnetic materials (FMs), where time-reversal symmetry ($\mathcal{T}$) is broken. However, antiferromagnetic materials (AFMs), characterized by zero net magnetization and symmetric magnetic configurations, have long been overlooked in AHE studies.[15,16] Despite this, AFMs offer unique advantages, such as ultrafast dynamics, robustness against external perturbations, and distinctive magnetotransport properties, making them highly attractive for future spintronic technologies.[17-20]

Recently, a new class of AFMs, known as altermagnets (AMs),[21-24] has attracted considerable attention due to their unconventional properties. These materials exhibit the alternating nature of the spin polarization in momentum space and nonrelativistic spin splitting occurring at generic $k$-points. Despite having zero net magnetization, AMs generate pronounced Hall signals, challenging the traditional understanding of the intrinsic AHE in AFMs.[25,26] Experimental studies on materials such as $RuO_2$ and $MnTe$,[21,27-29] which feature collinear compensated magnetic configurations, have provided direct evidence linking the observed AHE to the AM phase. A key factor driving this Hall response is the presence of finite Berry curvature. This reveals that even in centrosymmetric AFM crystals,[30] where symmetry would typically suppress the relativistic spin-orbit splitting, a significant intrinsic Hall response can still be induced, offering new insights into the AHE in AFM systems.[31-33]

While the Berry curvature is central to the linear intrinsic AHE, the real part of quantum geometry, known as the quantum metric,[7,8,10,34-36] plays an equally important role in transport phenomena,[37,38] particularly in the intrinsic nonlinear anomalous Hall effect (NAHE). In AFM systems with combined spatial inversion and time reversal symmetry ($\mathcal{PT}$), contributions from the Berry curvature and its dipole, which typically drive the intrinsic linear and extrinsic second-order nonlinear responses,[3,34,39] are suppressed due to symmetry constraints. Nevertheless, the quantum metric, allowed by these symmetries, becomes the dominant factor in generating an intrinsic second-order nonlinear response.[35] This $\tau$-independent NAHE results in a more complex and richer Hall behavior, as observed in $\mathcal{PT}$-symmetric AFM materials like $Mn_2Au$ and $CuMnAs$.[40-44] These findings highlight the crucial role of the quantum metric, demonstrating its significant impact on the Hall response, particularly in systems where the Berry curvature is excluded by symmetry.

Although significant progress has been made, few studies have managed to investigate these intrinsic Hall responses driven by different mechanisms within the same system. In addition, the intricate interplay between the real and imaginary parts of quantum geometry in shaping Hall responses in AFM systems remains poorly understood. In this work, we bridge this gap by demonstrating that careful manipulation of the interlayer magnetic coupling in a two-dimensional (2D) AFM bilayer can switch between AM and $\mathcal{PT}$-symmetric AFM phases. Using $CoPSe_3$ bilayers as a model system, we show that this transition enables us to realize two distinct Hall effects simultaneously: the intrinsic linear AHE in the AM phase, governed by the Berry curvature from generic $k$-points, and the intrinsic second-order NAHE in the $\mathcal{PT}$-symmetric AFM phase, driven by the quantum metric from high-symmetry $k$-points. Our work not only sheds light on the complex relationship between intrinsic linear and nonlinear Hall responses, but also demonstrates the potential of quantum geometry—particularly in AFMs, which have traditionally been overlooked in spintronics—to be flexibly tuned for novel device applications.

**Results.** Symmetry operations have distinct effects on the energy dispersion $E(k, \sigma)$. For example, spatial inversion ($\mathcal{P}$) reverses the wave vector ($k \rightarrow -k$) without affecting spin, while time reversal ($\mathcal{T}$) reverses both the wave vector and spin ($k, \sigma \rightarrow -k, -\sigma$). $\mathcal{PT}$-symmetry enforces spin degeneracy, preventing spin splitting across the entire Brillouin zone. Breaking $\mathcal{PT}$-symmetry is thus essential to induce spin splitting, which is a prerequisite for magnetoelectronic responses such as AHE. However, achieving spin splitting in AFM systems is challenging due to their inherent symmetry constraints. Notably, in 2D magnetic materials, interlayer magnetic coupling offers a route to manipulate symmetry properties and break $\mathcal{PT}$-symmetry.

In **Figure 1,** we take CoSeP$_3$ system as an example and illustrate how interlayer magnetic coupling can manipulate symmetry and break $\mathcal{PT}$-symmetry. The CoSeP$_3$ monolayer features a hexagonal crystal structure with a $D_{3d}$ point group in its nonmagnetic state. Each Co ion is situated within a distorted CoSe$_6$ octahedron, coordinated by six Se ions, with neighboring octahedra sharing edges (see **Figure S1**). As shown in **Figure 1a**, the Néel-type AFM ground state of CoSeP$_3$ monolayer retains $\mathcal{PT}$-symmetry. Consequently, the CoSeP$_3$ monolayer displays zero net magnetic moment and spin degeneracy throughout the Brillouin zone.

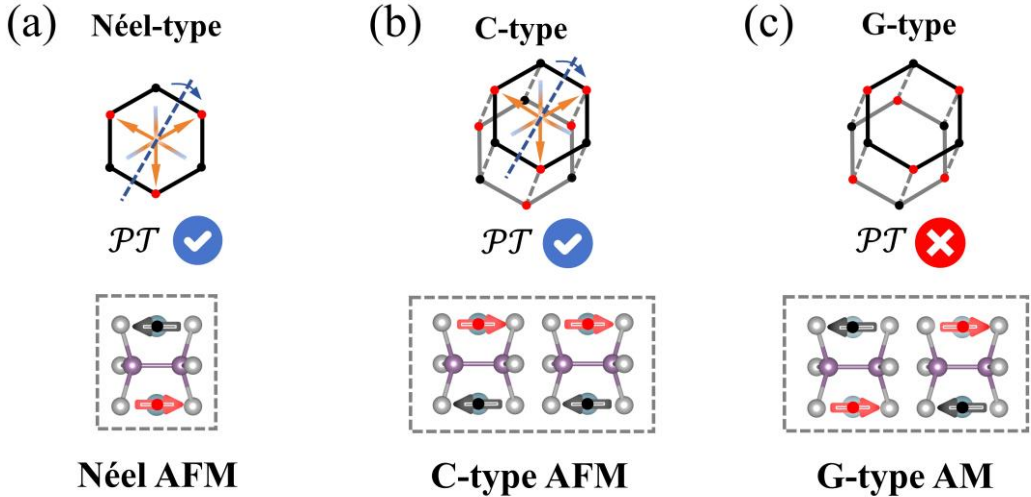

**Figure 1.** Crystal structures and magnetic configurations of 2D CoSeP$_3$ system. (a) The Néel-type AFM monolayer, (b) the C-type AFM bilayer, and (c) the G-type AM bilayer. Red and black points/arrows represent the magnetic moments of the magnetic ions.

In the AA-stacked bilayer (**Figures 1b** and **1c**), two magnetic configurations, namely C-type and G-type are considered. The C-type configuration is characterized by AFM intralayer coupling and FM interlayer coupling. This arrangement preserves $\mathcal{PT}$-symmetry and thus prohibits spin splitting. On the contrary, in the G-type configuration, both intralayer and interlayer couplings are AFM, leading to the breaking of $\mathcal{PT}$-symmetry. This symmetry breaking introduces spin splitting in the energy bands and gives rise to an AM phase. First-principles calculations further confirm that the G-type AM configuration is energetically more stable than the C-type AFM configuration in CoSeP$_3$ bilayers. Additional computational and structural details are provided in **Section I** and **II** of the **Supporting Information**.

To explore the intrinsic mechanism of magnetic coupling for the CoPSe$_3$ bilayer, we employ a spin-Heisenberg model,[45,46]

$$\hat{H}_{spin} = \sum_{n=1}^{3} J_n \sum_{\langle i,j \rangle} \hat{S}_i \cdot \hat{S}_j + J_c \sum_{\langle i,j \rangle} \hat{S}_i \cdot \hat{S}_j, \tag{1}$$

where $J_n$ ($n$ = 1, 2, 3) and $J_c$ represent the first to third nearest-neighbor of intralayer and the nearest-neighbor interlayer magnetic interactions, as illustrated in **Figures 2a** and **2b**, respectively. Here, $\hat{S}_i$ and $\hat{S}_j$ are spin operators corresponding to the magnetic moments at sites $i$ and $j$. By combining the spin-Heisenberg model with supercell mapping analysis,[45,46] the intralayer exchange parameters ($J_1 = -0.98$ meV, $J_2 = -0.01$ meV, $J_3 = 3.96$ meV) and the interlayer exchange parameter ($J_c = 0.08$ meV) are determined by modulating the spin orientation at sites $i$ and $j$ to eliminate undesired coefficients. The sign of $J$ provides insights into the nature of the interaction: positive $J$ indicates AFM coupling, while negative $J$ corresponds to FM coupling. Although $J_1$ and $J_2$ favor FM interactions, the strong AFM interactions of $J_3$ dominate the intralayer interactions. As a result, both the intralayer magnetic coupling (driven by $J_3$) and the interlayer coupling (determined by $J_c$) exhibit a clear preference for AFM ordering in CoPSe$_3$ bilayers.

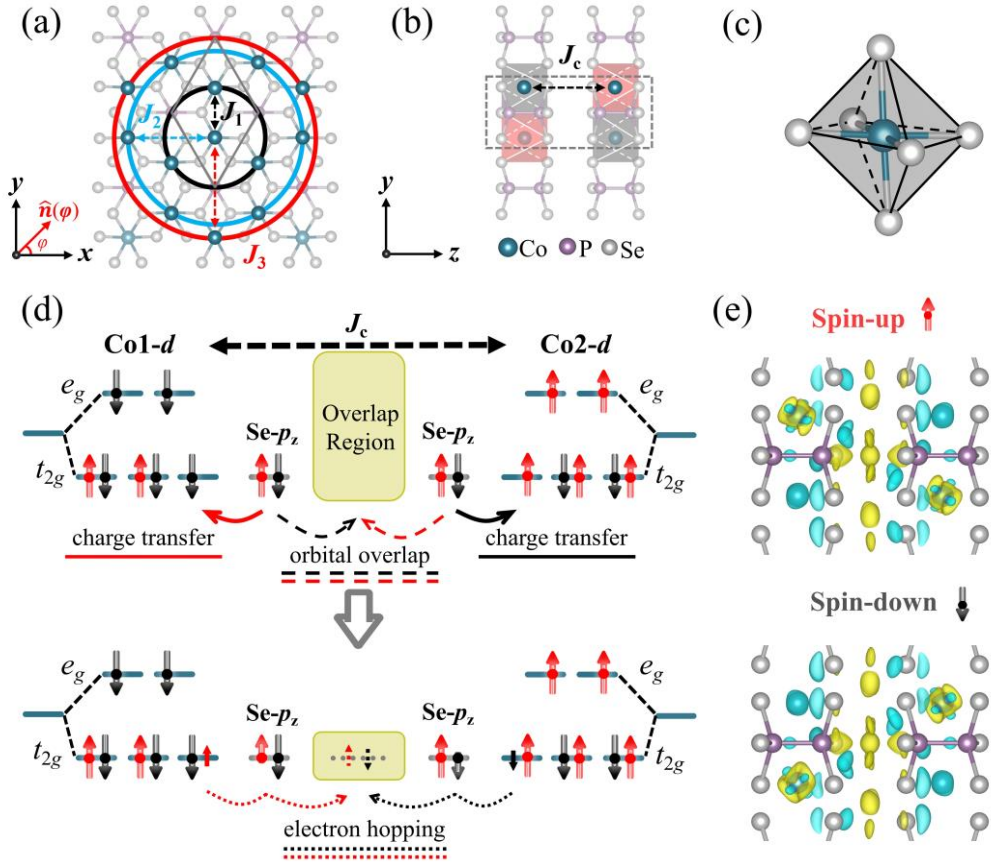

**Figure 2.** (a) Top view of CoPSe₃ bilayer, with the primitive cell outlined by a solid gray diamond. Nearest neighbors of the central Co atom are marked with black circles, second-nearest neighbors with blue circles, and third-nearest neighbors with red circles. The corresponding intralayer magnetic couplings are labeled as $J_1$, $J_2$, and $J_3$, respectively. The in-plane Néel vector is parameterized by $\hat{n}(\varphi)$ as a function of azimuth angle $\varphi$. (b) Side view of CoPSe₃ bilayer, with the interlayer magnetic coupling labeled as $J_c$. (c) Local structure of the distorted CoSe₆ octahedron. (d) Spin-exchange coupling mechanisms, with arrow lengths qualitatively indicating the number of spin-polarized electrons. Red and black curved arrows indicate charge transfer (solid line), orbital overlap (dashed line), and electron hopping (dotted lines). (e) Side view of the layer-resolved difference in charge density for spin-up and spin-down electrons in the ground-state AM configuration, with an isosurface value of $8 \times 10^{-5}$ e·(Bohr)⁻³. Yellow and blue contours indicate regions of charge accumulation and depletion, respectively.

These magnetic exchange results derived from the spin-Heisenberg model can be understood based on the superexchange mechanism.[47-49] For intralayer couplings, the bonding angle plays a critical role. When the Co-Se-Co bond angle is close to 90° (as observed in **Figure 2a** and **Figure S2a** for $J_1$), the superexchange contribution to AFM coupling is suppressed because electron transfer between cations requires the involvement of the same ligand orbital. In this case, the Se-4$p$ orbitals heading to the Co-$e_g$ orbitals mediate $e_g$-$e_g$ FM interactions (see **Figure S3a**). However, as the bonding angle deviates from 90°, the presence of the P side-group enables the degenerate $p$-orbitals along the extended Co-Se-P-Se-Co pathway (see **Figure 2a** and **Figure S2a** for $J_3$) to break the orthogonality constraint (see **Figure S3b**), thereby promoting $e_g$-$e_g$ AFM interactions. This mediation effect of the P side-group is further supported by the atom-resolved charge density illustrated in **Figure S3c**.

For interlayer exchange coupling, van der Waals (vdW) interactions form the basis for understanding the coupling mechanism. While interlayer interactions across the vdW gap are traditionally considered weak, recent studies have revealed their significant role in shaping and modifying electronic properties.[50] As illustrated in **Figures 2c** and **2d**, the Co-3$d$ orbitals experience $e_g$/$t_{2g}$ splitting under an octahedral crystal field. In the G-type AM configuration, substantial orbital overlap occurs between the Se-4$p$ orbitals in the interlayer P side-group and adjacent Se atoms, as evidenced by the layer-resolved differential charge density shown in **Figure 2e**. This orbital overlap, indicated by the dashed line in **Figure 2d**, facilitates an antiparallel spin charge transfer, represented by the red and black arrows. The transfer extends from the interface to the interlayer Co-3$d$ orbitals, highlighting the involvement of interfacial wave functions with both spin components. Notably, the nonpolarized overlap region minimizes Pauli repulsion, which in turn reduces the total energy of the CoPSe$_3$ bilayer. As a result, the Co-3$d$ orbitals remain partially occupied, which contributes to the interlayer AFM interactions.

In **Figures 3a** and **3b**, we display the electronic structures of C-type and G-type CoPSe$_3$ bilayers without incorporating relativistic spin-orbit coupling (SOC). Both configurations exhibit Kramers spin degeneracy along the high-symmetric $k$-path (Γ-K-M-K'-Γ). However, distinct differences emerge along the generic $k$-path (A-Γ-A'). In the C-type AFM configuration, the energy bands maintain spin degeneracy along both the high-symmetric and generic $k$-paths. In contrast, the G-type AM configuration exhibits pronounced spin splitting for a generic $k$, even in the absence of relativistic SOC. Furthermore, while spin alternates in momentum space, the net magnetization remains zero, as clearly demonstrated by its density of states (DOS).

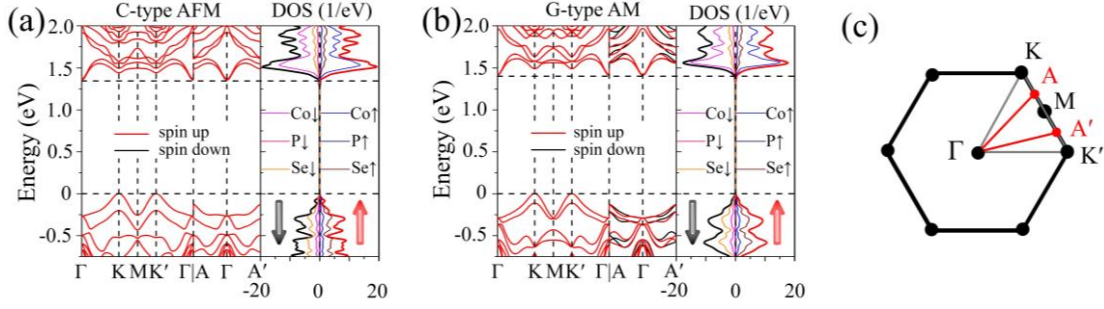

**Figure 3**. Band structures of CoPSe₃ bilayer of (a) C-type AFM configuration and (b) G-type AM configuration along high-symmetric *k*-path (Γ-K-M-K′-Γ) and generic *k*-path (A-Γ-A′). The corresponding atom-resolved DOS is attached to the side. (c) The corresponding first Brillouin zone of the hexagonal lattice with high-symmetry points labeled as Γ, M, K/K', and generic points A/A'. The letter A/A' marks the midpoint between the M and K/K' points.

To clarify the underlying mechanism, we carry out a symmetry analysis of the G-type AM phase. In its nonmagnetic state, the AA-stacked CoPSe₃ bilayer exhibits identity representation $\mathcal{E}$, inversion symmetry $\mathcal{P}$, two 3-fold rotation symmetry $\mathcal{C}_3$, three 2-fold rotation symmetry $\mathcal{C}_2$, two 6-fold rotation–reflection symmetry $\mathcal{S}_6$, and three mirror symmetry $\mathcal{M}$. Upon introducing collinear AFM order, both crystal and spin rotation symmetries should be considered, which remain decoupled in the absence of SOC. According to the pseudoscalar theory of spin,[51] the symmetry operations in the AM phase can be expressed as

$$\{\mathcal{E}, \mathcal{P}\} + \mathcal{U}\{3\mathcal{M}, 2\mathcal{C}_3, 3\mathcal{C}_2, 2\mathcal{S}_6\}. \qquad (2)$$

Here, $\mathcal{U}$ is an antisymmetric operation that reverses the signs of spin and magnetic moments without altering the wave vector $k$. The unitary symmetry operators $\mathcal{E}$ and $\mathcal{P}$ map each magnetic sublattice onto itself, while the other operators exchange opposite magnetic sublattices. The combined symmetry operations $\mathcal{U}\{\mathcal{M}, \mathcal{C}_3, \mathcal{C}_2, \mathcal{S}_6\}$ ensure spin degeneracy along high-symmetric *k*-paths (see **Section III** in **Supporting Information** for details). However, away from these paths, the distribution of the magnetization density around the Co sublattices is distorted (see **Figure 4a**). This distortion lifts the spin degeneracy, resulting in spin splitting in the Brillouin zone.

In **Figure 4b**, we illustrate the distribution of spin splitting, represented as $E_\uparrow - E_\downarrow$ for the valence band. The results show that the maximum amplitude of spin splitting occurs on both sides of the Γ-M path, with the value reaching up to 50 meV. In addition, the spin-split Fermi surfaces shown in **Figures 4c** and **4d** provide further evidence of spin polarization and symmetry breaking. These results demonstrate that the incorporation of spin degrees of freedom, together with interlayer magnetic coupling in vdW materials, can break symmetry and induce new physical phenomena within the homogeneous hexagonal system.

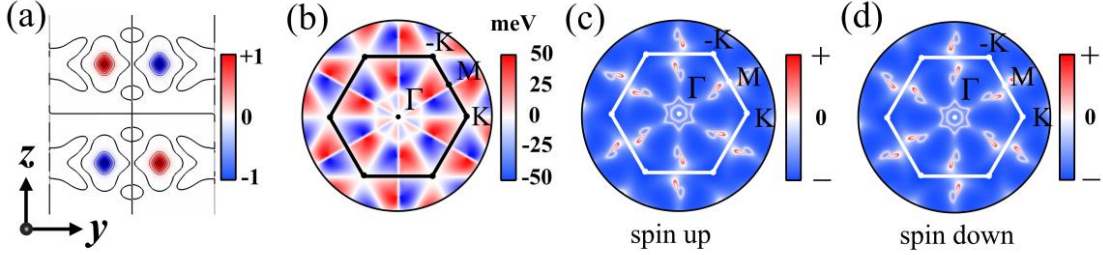

**Figure 4.** (a) Spin polarized charge densities of G-type AM CoPSe$_3$ bilayer along the center of the crystal plane. The isosurface is set as 0.03 e·(bohr)$^{-3}$. Red and blue colors indicate the different spin components. (b) Distribution of spin splitting ($E_\uparrow - E_\downarrow$) for the valence band without SOC. Corresponding Fermi surface of (c) spin-up and (d) spin-down states. The Fermi level is set at -0.29 eV below the VBM.

The emerging spin structures in G-type AM CoPSe$_3$ bilayer present opportunities for magnetoelectronic responses, enabling a straightforward method to achieve electrical control over magnetism. To investigate this response, we compute the anomalous Hall conductivity (AHC) using the Kubo formula,[1,2,52] which is derived by integrating the Berry curvature across the entire Brillouin zone as

$$\sigma_{\alpha\beta}^{AH} = -\frac{e^2}{\hbar} \int_{BZ} \frac{d\boldsymbol{k}}{(2\pi)^2} \sum_n f_{kn} \Omega_{\alpha\beta}^n(\boldsymbol{k}). \quad (3)$$

The $\alpha$ and $\beta$ indicate Cartesian coordinates. $f_{kn}$ is the Fermi-Dirac distribution function. $\Omega_{\alpha\beta}^n(\boldsymbol{k})$ is the band-projected Berry curvature, which is expressed as

$$\Omega_{\alpha\beta}^n(\boldsymbol{k}) = -2\mathrm{Im} \sum_{m \neq n} \frac{\langle n|\hat{v}_\alpha|m\rangle \langle m|\hat{v}_\beta|n\rangle}{\left(\varepsilon_{\boldsymbol{k}}^n - \varepsilon_{\boldsymbol{k}}^m\right)^2}. \quad (4)$$

Here, $|n\rangle$ denotes the eigenstates in band $n$ with eigenvalues $\varepsilon_{\boldsymbol{k}}^n$ and momentum $\boldsymbol{k}$.

$\hat{v}$ is the velocity operator.

As illustrated in **Figure 5b**, the AM CoPSe₃ exhibits strong anisotropy in the AHC with respect to the orientation of Néel vector. When the Néel vector is aligned along $\hat{n}(0°)$, the system retains its magnetic point group (MPG) 2/m.1, which imposes 2D nature and symmetry restrictions that prohibit the occurrence of the AHE. However, when the Néel vector is oriented along $\hat{n}(135°)$, the symmetry is lowered to MPG 2'/m', allowing a nonzero AHC component $\sigma_{xy}$. In this case, the AHC peaks at 46 S/cm at energy of -0.29 eV below the valence band maximum (VBM), as highlighted by the blue bubble in **Figure 5b**. The spin-resolved band structure of AM CoPSe₃ shown in **Figure 5a**, provides further insight into the origin of this AHC behavior. When the SOC is taken into account, the bilayer exhibits spin splitting with remarkable (anti)crossings at the Γ-point. Note that this crossing point highlighted by a color projection is not a coincidence but is determined by the alternating characteristic of spin polarization in momentum space.[21,53] Evidently, these crossings contribute significantly to the AHC peak observed in **Figure 5b**.

In **Figure 5c**, we show the *k*-dependent Berry curvature distribution associated with the peak in **Figure 5b**. The results reveal that the localized Berry curvature along the generic *k*-path directly contributes to the nonzero AHC, enabling the generation of a transverse current under an external electric field. The AHE is absent when the Néel vector is oriented along the *z*-axis (out of the plane), regardless of the magnetic phase. However, it emerges distinctly when the Néel vector lies within the intralayer plane of the AM phase, emphasizing the dependence of the AHE on both the magnetic phase and the Néel vector orientation. Relevant details regarding the symmetry restrictions and anisotropic AHE across different magnetic configurations are summarized in **Tables S1-S3**.

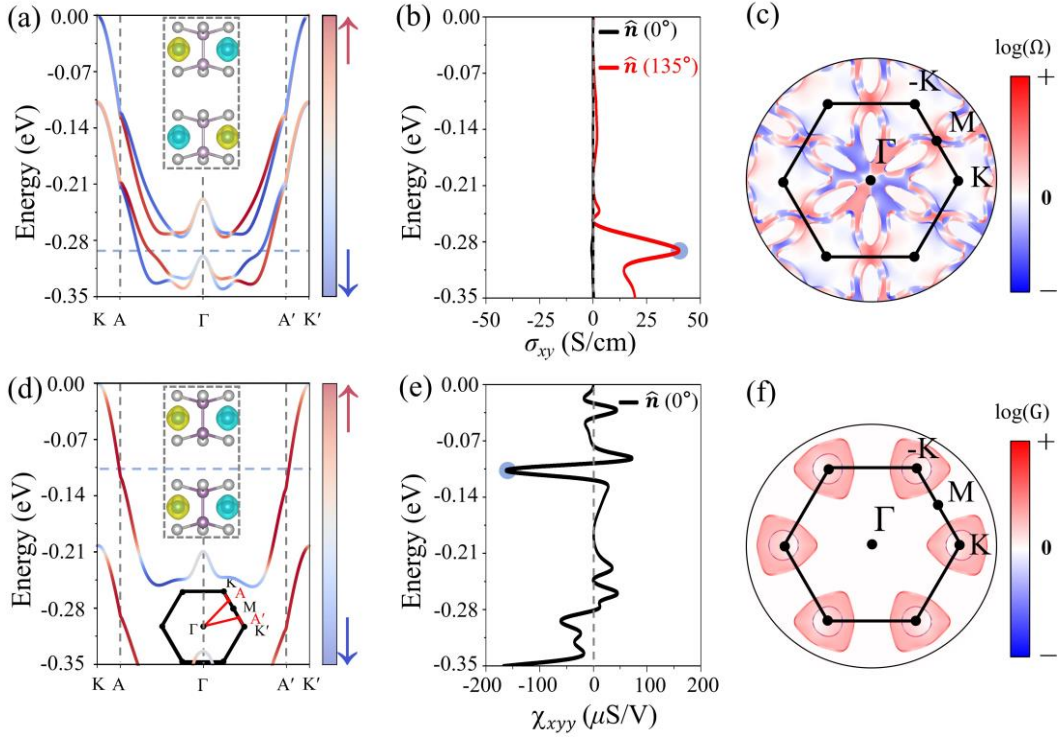

**Figure 5.** (a) Spin-resolved band structures in G-type AM CoPSe$_3$ bilayer with SOC. (b) The intrinsic AHC as a function of the Fermi energy for the G-type AM CoPSe$_3$ bilayer. The first peak is marked with a blue bubble, which is at -0.29 eV below the VBM. (c) The corresponding $k$-resolved Berry curvature in the 2D Brillouin zone. (d) Spin-resolved band structures in C-type AFM CoPSe$_3$ bilayer with SOC. (e) The second-order NAHE as a function of the Fermi energy for the C-type AFM CoPSe$_3$ bilayer. The peak highlighted by the blue bubble is located at -0.11 eV below the VBM. (f) The corresponding $k$-resolved BCP in the 2D Brillouin zone. A logarithmic transformation, as described in **Section I** of **Supporting Information**, is applied to both the Berry curvature and BCP. The VBM is set at the 0 eV.

As discussed above, the CoPSe$_3$ bilayer with the C-type AFM phase exhibits $\mathcal{PT}$-symmetry, which enforces spin degeneracy and suppresses the linear AHE (see **Figures 3a** and **5d**). Nevertheless, a higher-order anomalous Hall signal can still arise in this phase. According to the extended semiclassical theory,[35,43,54,55] the intrinsic second-order nonlinear anomalous Hall conductivity (NAHC) is given by

$$\chi_{\alpha\beta\gamma} = -\int_{\text{BZ}} \frac{d\boldsymbol{k}}{(2\pi)^2} \sum_n f_{kn} \left( \partial_\alpha G^n_{\beta\gamma}(\boldsymbol{k}) - \partial_\beta G^n_{\alpha\gamma}(\boldsymbol{k}) \right), \tag{5}$$

where $\alpha$, $\beta$, and $\gamma$ are Cartesian coordinates, $E$ is the component of the applied electric field, and $G^n$ is related to the local quantum geometry, known as the Berry connection polarizability (BCP) tensor. The band-projected BCP tensor is defined as[43,44]

$$G_{\alpha\beta}^n(\boldsymbol{k}) = 2Re \sum_{m \neq n} \frac{\mathcal{A}_\alpha^{nm}(\boldsymbol{k})\mathcal{A}_\beta^{nm}(\boldsymbol{k})}{\varepsilon_{\boldsymbol{k}}^n - \varepsilon_{\boldsymbol{k}}^m}. \qquad (6)$$

Here, $\mathcal{A}_\alpha^{nm} = \langle n|i\partial_{k_{\alpha^-}}|m\rangle$ is the interband Berry connection. The second-order NAHC as a function of Fermi energy is shown in **Figure 5e**. The peak highlighted by the blue bubble in the AFM CoPSe$_3$ bilayer ($\chi_{\alpha\beta\gamma} \approx 160\ \mu$S/V) is comparable to that of bulk antiferromagnet such as Mn$_2$Au (~150 $\mu$S/V) in reference [43]. This demonstrates the remarkable potential of the CoPSe$_3$ bilayer for nonlinear magnetoelectric applications.

In **Figure 5f**, we present the *k*-dependent distribution of BCP, corresponding to the peak marked by the blue bubble in **Figure 5e**. In contrast to the Berry curvature distribution associated with the linear AHC in the AM phase, the BCP in the C-type AFM phase is concentrated primarily near high-symmetric *k*-points. Specifically, the second-order NAHE is predominantly governed by the localized BCP at the K-point, highlighting its critical role in driving the observed nonlinear Hall response. Importantly, since the second-order NAHE holds the potential to significantly enhance detector responses and improve terahertz spectroscopic imaging, it presents a compelling opportunity for advancements in terahertz and infrared technologies.[56,57] Thus, the hexagonal CoPSe$_3$ bilayer, with tunable large anisotropic AHE and second-order NAHE via interlayer magnetic coupling, provides a promising platform for integrated devices.

In summary, we unveil a quantum geometric paradigm for realizing dual Hall effects in 2D CoPSe$_3$ bilayers via interlayer magnetic coupling. The AM phase generates a robust anisotropic AHE governed by Berry curvature hotspots, while the $\mathcal{PT}$-symmetric AFM phase exhibits an intrinsic second-order NAHE dominated by quantum metric contributions at high-symmetric *k*-points. The reversible transition between these phases, mediated by interlayer exchange interactions, underscores the critical role of quantum geometry in linking band structures to transport phenomena. Beyond demonstrating the coexistence of linear and nonlinear Hall responses, our findings reveal how symmetry engineering in layered antiferromagnets can selectively activate distinct quantum geometric mechanisms. These conclusions can be extended to few-layer systems, as detailed in **Section V** of the **Supporting Information**. The integration of AHE and NAHE in a single material system opens avenues for multifunctional device architectures, where quantum geometry serves as a cornerstone for emergent magnetoelectric functionalities.

**Notes**

The authors declare no competing financial interest.

■ **ACKNOWLEDGMENTS**

We are grateful for the Shenzhen Natural Science Fund (the Stable Support Plan Program 20231121110218001), the Guangdong Basic and Applied Basic Research Foundation (2022A1515012006), and the National Natural Science Foundation of China (Grant Nos. 12104321 and 12034014).

# Quantum Geometric Engineering of Dual Hall Effects in 2D Antiferromagnetic Bilayers via Interlayer Magnetic Coupling

Zhenning Sun[1], Tao Wang[1], Hao Jin[1]*, Xinru Li[2], Yadong Wei[1]*, Jian Wang[1,3]

[1]*College of Physics and Optoelectronic Engineering, Shenzhen University, Shenzhen 518060, P. R. China*

[2]*School of Physics, State Key Laboratory of Crystal Materials, Shandong University, Shandanan Street 27, Jinan 250100, China*

[3]*Department of Physics, The University of Hong Kong, Pokfulam Road, Hong Kong 999077, China*

# CONTENTS

## I. Computational Methods

First-principles calculations employing the projector augmented wave (PAW) method are performed by the DFT software Vienna ab initio simulation package (VASP).[1,2] The exchange-correlation function is approximated by the generalized gradient approximation (GGA) within the Perdew–Burke–Ernzerhof (PBE) form.[3,4] The criteria of energy and force are set to $10^6$ eV and $10^{-2}$ eV·Å$^{-1}$, respectively, with a plane wave cutoff energy of 500 eV for the guarantee of convergence. The $k$-point meshes are set to $12 \times 12 \times 1$. An effective Hubbard $U_{eff}$ is applied to the Co-$d$ orbitals by the GGA + $U$ ($U = 4.0$ eV) method to include the strong Coulomb interaction on the $3d$ orbitals.[5,6] A vacuum layer of 20 Å is added to eliminate uncorrelated interlayer interaction in the periodic calculations. The semi-empirical dispersion correction (DFT-D3) method is used to capture nonbonding van der Waals (vdW) interactions.[7,8] Magnetic exchange parameters are achieved through the mapping analysis.[9] The tight-binding Hamiltonian that considers Co-$3d$, P-$3p$, and Se-$4p$ orbitals is constructed through the maximally localized Wannier functions (MLWFs) generated by *wannier*90 code.[10]

Berry curvature $\Omega$ and Berry connection polarizability (BCP) $G$ are intrinsically related to the experimentally detected quantity (current $j$) associated with the AHE and the second-order NAHE. The anomalous Hall conductivity (AHC) $\sigma_{\alpha\beta}^{AH}$ and second-order intrinsic nonlinear anomalous Hall conductivity (NAHC) tensor $\chi_{\alpha\beta\gamma}$ are defined as：

$$\begin{cases} j_\alpha^{AHE} = \sigma_{\alpha\beta}^{AH} E_\beta \\ j_\alpha^{NAHE} = \chi_{\alpha\beta\gamma} E_\beta E_\gamma \end{cases} \tag{1}$$

Here, $\alpha$, $\beta$, and $\gamma$ are Cartesian coordinates. $E$ is the component of the applied electric field. The AHC is defined as $\sigma_{\alpha\beta}^{AH} = -\frac{e^2}{\hbar} \int_{BZ} \frac{d\boldsymbol{k}}{(2\pi)^2} \Omega_{\alpha\beta}(\boldsymbol{k})$. The second-order NAHC is defined as $\chi_{\alpha\beta\gamma} = -\int_{BZ} \frac{d\boldsymbol{k}}{(2\pi)^2} \sum_n (\partial_\alpha G_{\beta\gamma}^n - \partial_\beta G_{\alpha\gamma}^n) f_0$. $f_0$ is the Fermi-Dirac distribution function.

The Berry curvature $\Omega$ and BCP $G$ calculated by a dense $k$-mesh of $600 \times 600 \times 1$ exhibit a more localized representation compared to other materials. To promote visualization, we apply a logarithmic transformation to the results:[11,12]

$$\log(X) = \begin{cases} sgn(X)\lg(X), & |X| > 10; \\ X/10, & |X| \leq 10. \end{cases} \qquad (2)$$

Here, $sgn(X)$ is the signum function return back to the positive or negative sign of the original X ($X = \Omega$ or $G$).

## II. Geometric Information and Superexchange Process of CoPSe₃ Bilayer

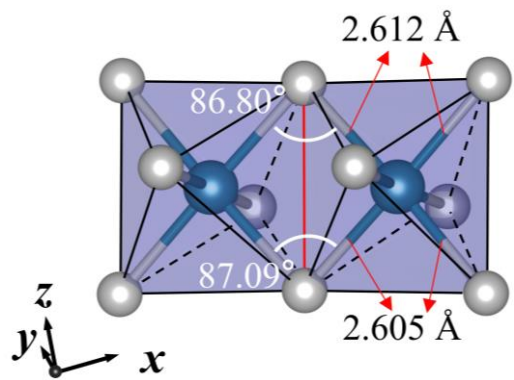

**FIG. S1.** Local structure of neighboring CoSe₆ octahedra. The solid red line is the boundary of the neighboring octahedra.

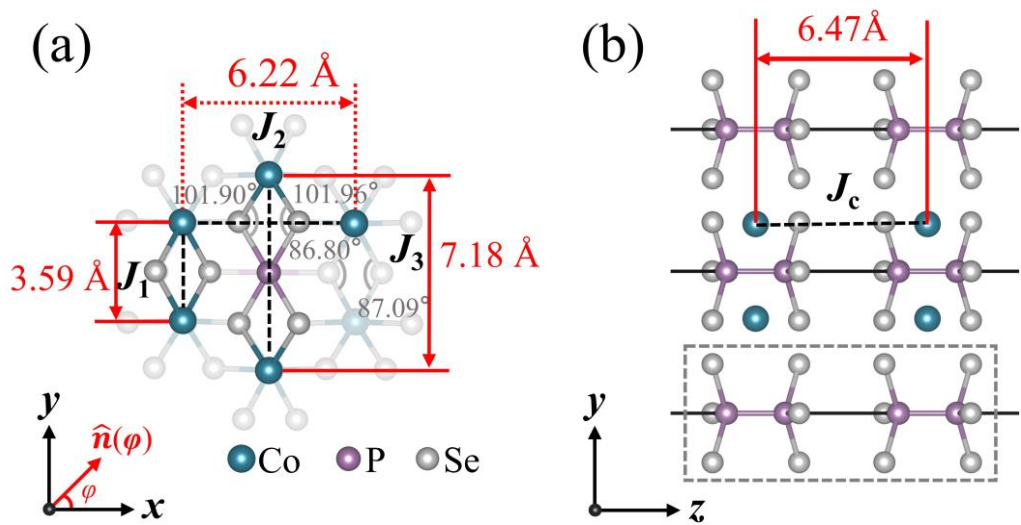

**FIG. S2.** (a) Geometric information among magnetic atoms with intralayer magnetic coupling coefficients $J_1$, $J_2$, and $J_3$. (b) Geometric information among magnetic atoms with interlayer magnetic coupling coefficient $J_c$.

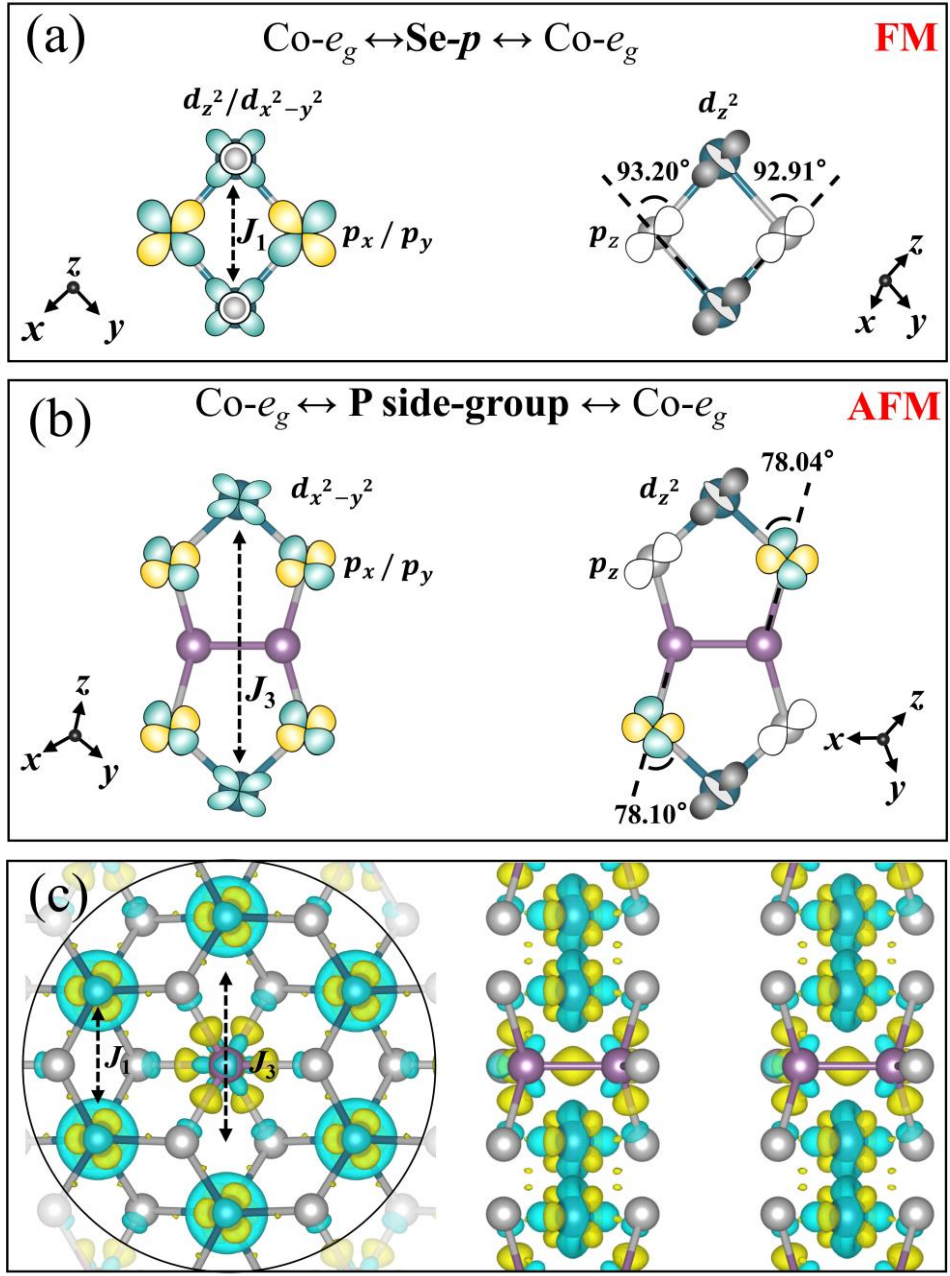

**FIG. S3.** (a) Qualitative illustration of possible orbital orientation along the Co-Se-Co path for the nearest neighbors interaction $J_1$. (b) Possible orbital orientation along the Co-Se-P-Se-Co path for the third-nearest neighbors interaction $J_3$. For visual unity, the views of geometries for the same path are consistent while the reference coordinates of ligand are slightly revised. (c) Top and side views of the atom-resolved difference in charge density with an isosurface value of 0.015 e·(bohr)$^{-3}$. Yellow and blue contours indicate charge accumulation and depletion, respectively.

## III. Symmetry Analysis

A detailed analysis of symmetry for spin splitting is as follows. With a space group of $P\bar{3}1m$ (no. 162) and a point group of $D_{3d}$, the AA-stacked CoPSe₃ bilayer includes an identity representation $\mathcal{E}$, an inversion symmetry $\mathcal{P}$, two 3-fold rotation symmetry $\mathcal{C}_3$, three 2-fold rotation symmetry $\mathcal{C}_2$, two 6-fold rotation–reflection symmetry $\mathcal{S}_6$, and three mirror symmetry $\mathcal{M}$. Following the pseudoscalar theory of spin, the symmetry operations for the CoPSe₃ with the altermagnetic (AM) phase can be described as

$$\{\mathcal{E}, \mathcal{P}\} + \mathcal{U}\{3\mathcal{M}, 2\mathcal{C}_3, 3\mathcal{C}_2, 2\mathcal{S}_6\}. \tag{1}$$

Here, $\mathcal{U}$ is an antisymmetric operation that reverses the signs of spin and magnetic moments without affecting the wave vector $k$. The unitary symmetry operators $\mathcal{E}$ and $\mathcal{P}$ map each magnetic sublattice onto itself, while others map opposite magnetic sublattices onto each other. For example, applying the combined symmetry operation $\mathcal{U}\mathcal{M}_x$ on the system yields

$$\mathcal{U}\mathcal{M}_x E\left(k_x, k_y, \uparrow\right) = E\left(-k_x, k_y, \downarrow\right). \tag{2}$$

Here, the symbol $\uparrow$ and $\downarrow$ represent different components of spin. The equation states that different spins in band energies degenerate at $k_x = 0$. Combining this with Bloch theorem yields

$$E\left(\pm\pi, k_y, \uparrow\right) = E\left(\pm\pi, k_y, \downarrow\right) \tag{3}$$

Consequently, Kramer spin degeneracy is maintained in energy dispersion along high-symmetry $k$ paths with $k_x = n\pi$. Similarly, the combined symmetry operation $\mathcal{U}\mathcal{M}_y$ guarantees the spin degeneracy of high-symmetry $k$ paths with $k_y = m\pi$ in the energy dispersion, when the combined symmetry operation $\mathcal{U}\{\mathcal{C}_3, \mathcal{S}_6\}$ guarantees spin degeneracy on the paths of $k_x = n\pi/6$ and $k_y = m\pi/6$. Thus, spin splitting of the high-symmetry paths of the hexagonal CoPSe₃ bilayer is symmetry forbidden. Away from these high symmetry paths, the distortion of the magnetization density within the layers breaks the $\mathcal{T}\mathcal{C}_2$ symmetry of the interlayer and therefore allow for the spin splitting.

In the AM phase, the CoPSe₃ bilayer exhibits a strong anisotropy of the AHC. When the Néel vector is oriented along the $\hat{n}(135°)$, the magnetic point group (MPG) retains 2'/m' (no. 5.5.16). The symmetry operators are described as

$$\{\mathcal{E}, \mathcal{P}\} + \mathcal{T}\{\mathcal{M}_y, \mathcal{C}_{2y}\}. \tag{4}$$

The spatial symmetries are coupled with the spin-space symmetry. Therefore, there exists a nonzero component $\sigma_{xy}$ of the AHC in the CoPSe₃ bilayer. However, when the Néel vector is oriented along $\hat{n}(0°)$, the MPG becomes 2/m.1 (no 5.1.12). The symmetry operator is described as

$$\{\mathcal{E}, \mathcal{P}, \mathcal{M}_y, \mathcal{C}_{2y}\}. \tag{5}$$

In this case, the AHE is prohibited as the symmetry constraints in 2D system.

## IV. Anisotropy of the Abnormal Hall Effect (AHE) and the Second-order Intrinsic Nonlinear Anomalous Hall Effect (NAHE)

In Tables S1, S2, and S3, we list the corresponding magnetic symmetries and indicate the presence (√) or absence (×) of the AHE and the NAHE in both CoPSe$_3$ monolayer and bilayer, respectively. Note that the spin orientations for in-plane cases (*xy*-plane) of different magnetic phases are labeled as $\hat{n}(\varphi)$, when the out-of-pane (along the *z*-axis) case is described separately.

**Table S1.** The MPGs and the presence or absence of AHE and NAHE for Néel-type AFM CoPSe$_3$ monolayer.

| Néel-type AFM | | | | |
|---|---|---|---|---|
| Monolayer | Néel vectors $\hat{n}(\varphi)$ | MPGs | Linear AHE | Second-order NAHE |
| state1 | $\hat{n}(0°)$ | 2′/m | × | √ |
| state2 | $\hat{n}(90°)$ | 2′/m | × | √ |
| state3 | $\hat{n}(135°)$ | 2/m′ | × | √ |
| state4 | out-of-plane | -3′m | × | × |

**Table S2.** The MPGs and the presence or absence of AHE and NAHE for C-type AFM CoPSe$_3$ bilayer.

| C-type AFM | | | | |
|---|---|---|---|---|
| Bilayer | Néel vectors $\hat{n}(\varphi)$ | MPGs | Linear AHE | Second-order NAHE |
| state1 | $\hat{n}(0°)$ | 2′m | × | √ |
| state2 | $\hat{n}(90°)$ | 2′m | × | √ |
| state3 | $\hat{n}(135°)$ | 2/m′ | × | √ |
| state4 | out-of-plane | -3′m | × | × |

**Table S3.** The MPGs and the presence or absence of AHE and NAHE for G-type AM CoPSe$_3$ bilayer.

| Bilayer | **G-type AM** Néel vectors $\hat{n}(\varphi)$ | MPGs | Linear AHE | Second-order NAHE |
|---|---|---|---|---|
| state1 | $\hat{n}(0°)$ | 2/m.1 | × | × |
| state2 | $\hat{n}(90°)$ | 2/m.1 | × | × |
| state3 | $\hat{n}(135°)$ | $2'/m'$ | √ | × |
| state4 | out-of-plane | -3m.1 | × | × |

It is clear that the linear AHE exists exclusively in the AM phase and exhibits a pronounced anisotropy. In AA-stacked CoPSe$_3$ bilayers, the presence or absence of linear AHE is closely related to the symmetry breaking induced by the magnetic arrangement and the resulting nonrelativistic spin splitting. Thus, the existence of linear AHE depends on the orientation of Néel vector. Consequently, control of the linear AHE can be effectively achieved through combined manipulation of both crystal symmetry and magnetic configuration. For NAHE, anisotropy is observed in both monolayers and bilayers. Due to the symmetry restriction, both AHE and NAHE are prohibited with out-of-plane Néel vectors.

## V. Spin Splitting in Trilayer System

The structure of the AA-stacked CoPSe₃ trilayer is illustrated in **Figures S4a** and **S4b**. The top view of CoPSe₃ trilayer resembles that of the bilayer while the side view is similar to bilayer for the same stacking pattern. Through the symmetry analysis of three different interlayer magnetic configurations presented in **Figure S4c**, we find that both the $\mathcal{PT}$-symmetry in the C-type AFM and the G-type AFM configurations are preserved, while the $\mathcal{PT}$-symmetry is broken in the hybrid magnetic configuration.

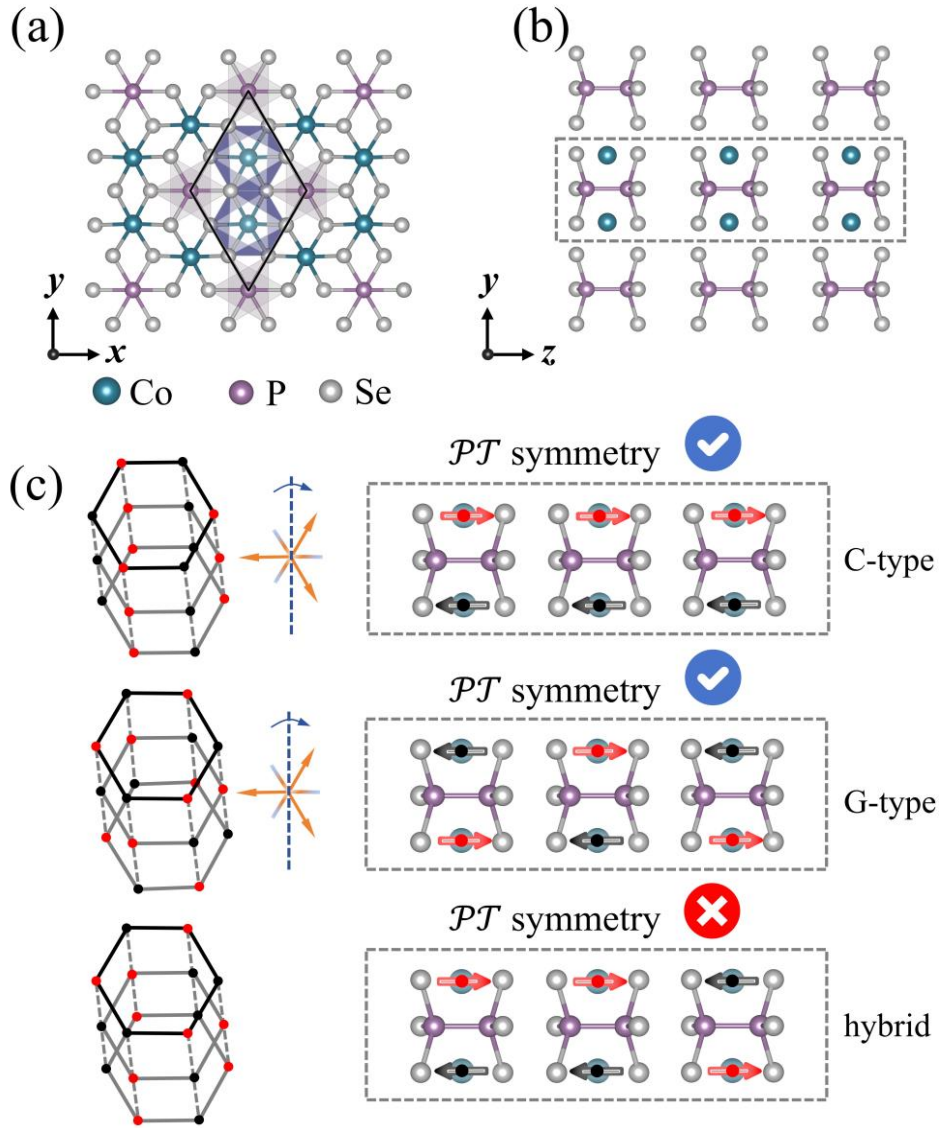

**FIG. S4.** Crystal and magnetic configurations of CoPSe₃ trilayer. (a) Top and (b) side views of CoPSe₃ trilayers with solid black diamonds labeling the primitive cell. (c) Spin-dependent $\mathcal{PT}$ symmetry in trilayer, where the red and blue points/arrows indicate the magnetic moments of Co ions.

To further verify the electronic structure, we show the band structures of the three magnetic configurations corresponding to the generic *k*-paths in **Figure S5**. Obviously, nonrelativistic spin splitting occurs in the symmetry-broken interlayer hybrid magnetic states. These findings highlight the potential occurrence of nonrelativistic spin splitting in multilayer hexagonal systems.

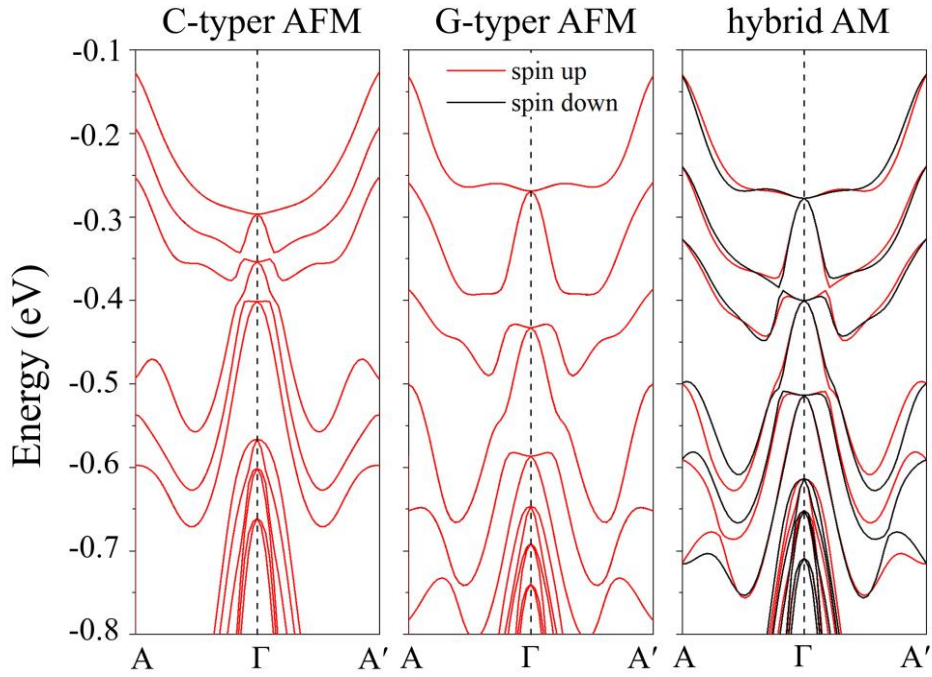

**FIG. S5.** Band structures of CoPSe$_3$ trilayer with different magnetic configurations.